\documentclass{PoS}

\title{Finite temperature QCD at fixed Q with overlap fermions}

\ShortTitle{FT QCD at fixed Q with overlap fermions}

\author{JLQCD Collaboration: \speaker{Guido Cossu}$^a$\thanks{E-mail: cossu@post.kek.jp}, Sinya Aoki$^b$, Shoji Hashimoto$^{a,c}$, Takashi Kaneko$^{a,c}$, Hideo Matsufuru$^a$, Jun-ichi Noaki$^a$, Eigo Shintani$^d$\\
\\
\llap{$^a$}Theory Center, IPNS, High Energy Accelerator Research Organization
  (KEK), Tsukuba 305-0801, Japan\\
\llap{$^b$}Graduate School of Pure and Applied Sciences, University of Tsukuba, Tsukuba 305-8571, Japan\\
\llap{$^c$}School of High Energy Accelerator Science, The Graduate University for
Advanced Studies (Sokendai), Tsukuba 305-0801, Japan\\
\llap{$^d$}RIKEN-BNL Research Center, Upton, NY 11973-5000, USA}

\abstract{We present some preliminary results of the project on finite temperature QCD with overlap fermions at KEK. We performed a series of simulations to assess the effects of fixing the topological sector at finite temperature and we will show the first calculations of topological susceptibility and meson masses for quenched and full QCD.}

\FullConference{The XXVIII International Symposium on Lattice Field Theory, Lattice2010\\
		June 14-19, 2010\\
		Villasimius, Italy}

\begin{document}

\section{Introduction and motivation}

Among several features of QCD the chiral symmetry breaking is one of the most interesting ones. At low temperature this symmetry is spontaneously broken and the vacuum develops a quark anti-quark  condensate, $\langle \bar q q \rangle \neq 0$. Massless Nambu-Goldstone (NG) bosons should appear in the spectrum of the massless theory. In real QCD, of course, the light quark masses explicitly break the chiral symmetry, giving a small mass to the NG bosons that have been identified as the 8 lightest mesons (pions, kaons, $\eta$). Classically, the pattern of chiral symmetry breaking is the following ($N_f$ being the number of light quarks)\footnote{$U(1)_V$ gives the conserved baryon number and $SU(N_f)_V$ is only softly broken by the small quark mass difference.}:
\begin{equation}
  SU(N_f)_V \times SU(N_f)_A \times U(1)_V \times U(1)_A \rightarrow SU(N_f)_V \times U(1)_V  
\end{equation}
that should actually give 9 NG bosons, but the ninth particle is absent in the spectrum.  This apparent problem was solved noticing that the flavor-singlet axial $U(1)_A$ rotation is no more a symmetry at the quantum level (even in the massless limit); it is anomalous \cite{'tHooft:1976up} due to the presence of instanton-like configurations. A direct effect of the anomaly is the large splitting in the mass of flavor-singlet and non-singlet pseudoscalar mesons  (see Witten-Veneziano formula \cite{Witten:1979vv,Veneziano:1979ec}).

While at zero temperature the physics is quite clear, at finite temperature still there is no definite answer to the question if axial $U(1)_A$ symmetry is restored or not. If it is restored, an interesting problem is to establish if this happens at the same critical temperature of chiral symmetry restoration. This would have relevant effects on the pattern of symmetry breaking and so on the critical exponents of the phase transition \cite{Butti:2003nu}.

By semiclassical calculations of dilute instanton gas at very high temperature $T \gg T_c$, we expect a strong suppression but not an exact restoration of the $U(1)_A$ symmetry.
The most advanced lattice result in this context is the one by Vranas \cite{Vranas:1999dg}, where, using domain wall fermions, he concluded that just above chiral phase transition the $U(1)_A$ symmetry remains broken but by a very small amount. It is an open question what is the effect that may have to the order of the transition. 

Our aim is to study the fate of chiral and $U(1)_A$ symmetry (and the mass of particles) at finite temperature around and above the phase transition using the fermionic action that retains the maximal amount of chirality on the lattice, {\it i.e.} the overlap formulation \cite{Luscher:1998pqa}.
The JLQCD and TWQCD collaborations have performed large scale QCD simulations using the overlap action \cite{Kaneko:2006pa}. All simulations were done at zero temperature and we investigated the chiral behavior of spectra, low energy constants, chiral condensate and topological susceptibility \cite{Aoki:2007pw,Noaki:2008iy,Aoki:2009qn,Fukaya:2009fh}. A non-zero topological susceptibility, indicating anomalous breaking of $U(1)_A$ symmetry, was clearly observed in those simulations at $T=0$ and also the linear dependence with sea quark mass was obtained, as predicted by chiral perturbation theory. 

In order to simulate QCD using HMC with dynamical overlap fermions, fixing the topological sector was crucial because allowing for topology change would be extremely expensive \cite{Fukaya:2006vs}. To run a simulation at fixed topological charge $Q$ it was introduced an irrelevant term in the action \cite{Fukaya:2006vs}, that suppresses the occurrence of zero eigenvalues of the hermitian Wilson-Dirac operator that have to be crossed in order to change topological sector. Fixing topology, of course, creates a bias in physical results that must be corrected. A full theory describing the effects of working at fixed $Q$ was developed \cite{Aoki:2007ka}: the effects at zero temperature are understood, under control and $O(1/V)$. We will discuss with more details the subject in the following section.

In order to obtain reliable results at finite temperature we need to check whether the same methods used at  zero temperature to correct for fixed topology effects work even in this case. So, we started with some exploratory studies using quenched theory but fixing topology in order to compare with previous results in the literature. We measured the topological susceptibility  and several meson correlators at finite temperature. We will report the results of these simulations and the preliminary results in full QCD to investigate $U(1)_A$ restoration.





\section{Simulations and results}
Before discussing the results, let us briefly describe the methods used to measure correlators and the topological susceptibility at fixed topology. Detailed derivation of equations can be found in \cite{Aoki:2007ka}.

By using a saddle point expansion of the QCD partition function in a finite volume we can derive an expression for $Z_Q$, partition function at fixed topology ($V$ is the 4-volume):
\begin{equation}
Z_Q = \frac{1}{\sqrt{2\pi \chi_t V}}\exp\Bigl[-\frac{Q^2}{2 \chi_t V}\Bigr]\Bigl[1- \frac{c_4}{8V\chi_t^2}+ O\Bigl(\frac{1}{V^2}\Bigr)\Bigr],
\label{eq:FixedQ}
\end{equation}
a gaussian distribution for topological charge that can be used to show that:
\begin{equation}
\lim_{|x| \rightarrow \infty} \langle \rho(x) \rho(0) \rangle = \frac{1}{V}\Bigl(\frac{Q^2}{V}-\chi_t - \frac{c_4}{2\chi_t V}\Bigr) + O(V^{-3}).
\label{eq:TopChargeCorr}
\end{equation}
which implies that the topological susceptibility can be extracted from a long range correlation of the topological charge density $\rho(x)$. At first order in $1/V$ we can ignore the contribution of the $c_4$ term, and check later the consistency of the assumption.

Using the overlap operator we can define an object that has the same properties as the topological charge density:
\begin{equation}
\rho_m(x) = m \, {\rm tr} [\gamma_5 (D_c + m)_{x,x}^{-1}],  
\end{equation}
where $(D_c+m)^{-1}$ is the valence quark propagator constructed using the chirally symmetric overlap operator.  An alternative way is to consider the pseudoscalar isosinglet $\eta'$ correlator, whose disconnected part is equal to $\langle \rho_m(x) \rho_m(0) \rangle$ at large distances and couples only to the fast decaying $\eta'$ state, making it a better choice in order to estimate the long distance limit.

We measured the connected and disconnected part of pseudoscalar correlators at long distance to extract $\chi_t$ \cite{Aoki:2007pw}. Since we are working at finite temperature the correlators are measured and averaged over spatial directions. We reconstructed the correlators by using the first 50 eigenvectors of the overlap operator assuming low mode dominance. For example, the connected scalar correlator is given by  
\begin{equation}
C(x,y)_N = {\rm Tr} \sum_{ij}^N \frac{\psi_{\lambda_i}(x)\psi^\dagger_{\lambda_i}(y)}{i\lambda_i+m} \frac{\psi_{\lambda_j}(y)\psi^\dagger_{\lambda_j}(x)}{i\lambda_j+m}  
\end{equation}
and similar expression of the pseudoscalar (just $i\lambda_i \rightarrow -i\lambda_i$). We checked that the saturation with 50 eigenmodes is sufficiently accurate for the infrared behavior.

\subsection{Pure gauge simulations}

By introducing the topology fixing term in pure gauge simulations we can check if, even at finite temperature, we can reconstruct topological susceptibility using the method described above by comparing with the literature.

The setup is the following: Iwasaki action + topology fixing term at temperatures ranging from $[0.8,1.3]T_c$ on two different volumes $16^3\times 6$ and $24^3\times 6$. The critical point was estimated to be at $\beta_c = 2.445$ by inspecting the Polyakov loop.

\begin{figure}
\center
\includegraphics[width=0.65\textwidth,clip=1]{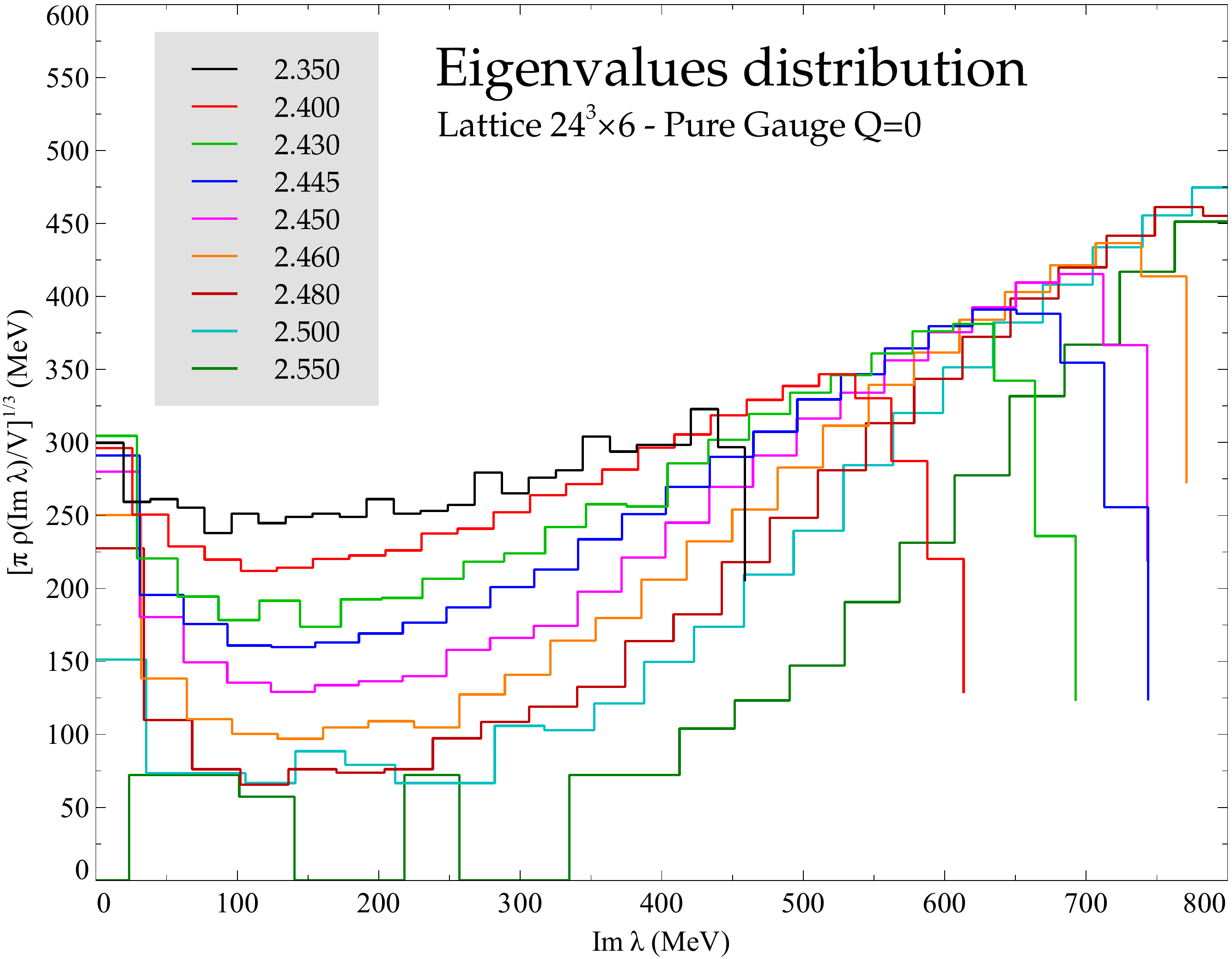}
\caption{Spectral density of the overlap-Dirac operator at finite temperature on the $24^3\times 6$ quenched lattice.}
\label{fig:Eigenvals24}
\end{figure}

We first check whether the eigenvalue distribution behaves as expected. The typical distribution is shown in Figure \ref{fig:Eigenvals24}. We do not find any discrepancy with previous results (e.g. \cite{Edwards:1999zm}). The presence of a peak for small eigenvalues in the high temperature side ($T>T_c$) was confirmed. In \cite{Edwards:1999zm} these modes are associated with the presence of dilute gas of instantons-anti instantons.

\begin{figure}
\center
\includegraphics[width=0.65\textwidth,clip=1]{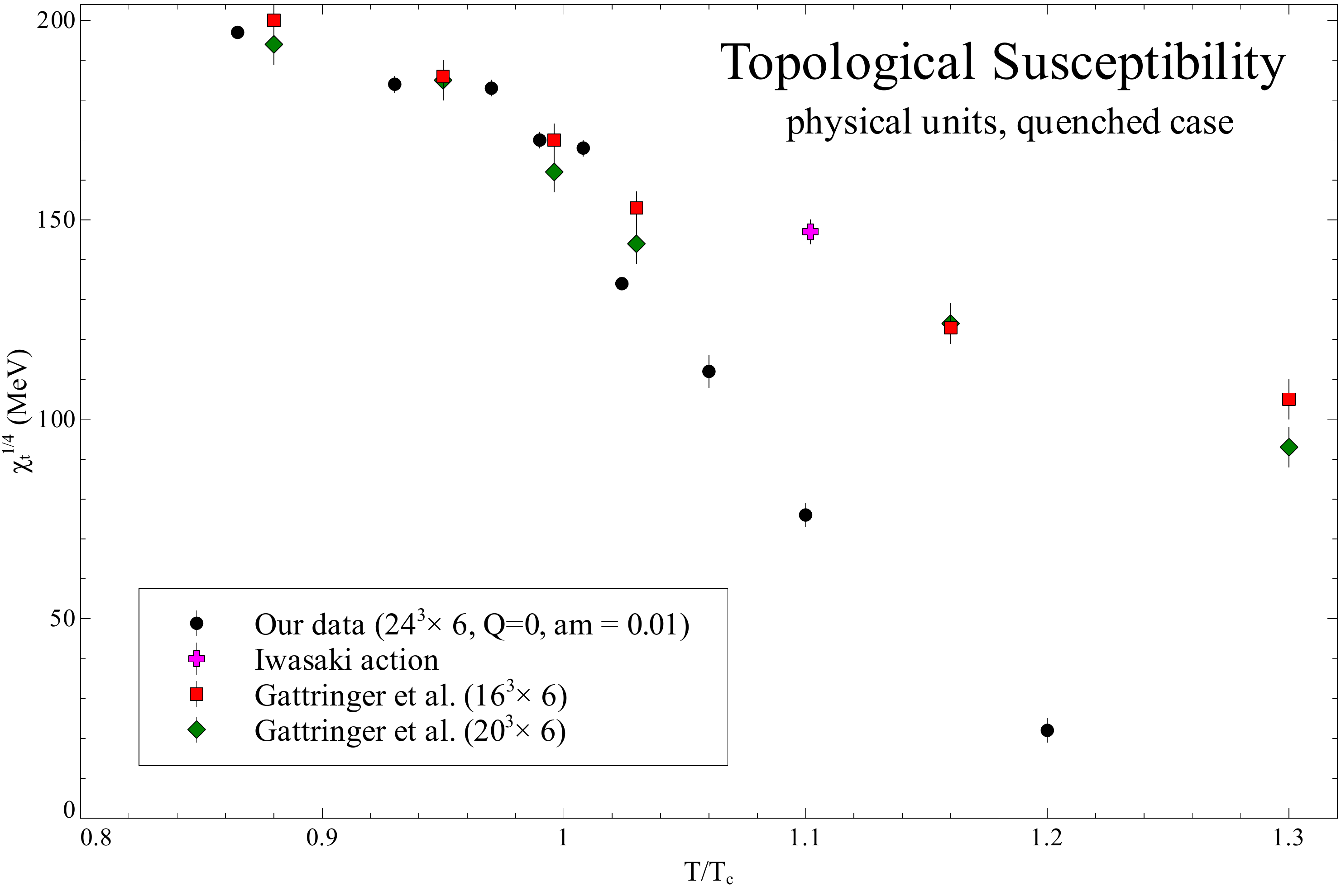}
\caption{Comparison of topological susceptibility results with \cite{Gattringer:2002mr}, lattice $24^3\times 6$}
\label{fig:TopSusceptComparison}
\end{figure}

The most interesting result at this stage is the behavior of the topological susceptibility at finite temperature in comparison with the results of Gattringer {\it et al.} \cite{Gattringer:2002mr}, shown in Figure \ref{fig:TopSusceptComparison}. The asymptotic value for the disconnected correlator (see equation (\ref{eq:TopChargeCorr}) ) was estimated using a joint fit of the connected and disconnected parts and assuming a double pole form for the last one in the quenched theory. The decay is dominated by pionic states for both of them. Then we use (\ref{eq:TopChargeCorr}) to extract $\chi_t$ assuming that the $c_4$ term is negligible. 

In \cite{Gattringer:2002mr} the topological susceptibility was measured using the index theorem by just counting the number of zero modes of an approximated version of the overlap operator, which is a clean definition without the ambiguities due to cooling. By also performing a simulation without the topology fixing term, we checked their results also with our exact overlap operator, finding no significant deviations within errors (cross symbol in Figure \ref{fig:TopSusceptComparison}, temperature $T/T_c=1.1$).

Below the transition temperature there is agreement between the two sets of data, showing that the method works very well at least until $T_c$. Above the critical temperature our results are systematically lower than the reference ones. We are currently investigating the source of this discrepancy which could be traced back in the assumptions leading to (\ref{eq:FixedQ}) or, the $c_4$ term could be non negligible in this regime. We are trying to estimate this quantity but still we are getting too large error bars for the four-point spatial correlators.  

Another possibility to check the validity of (\ref{eq:TopChargeCorr}) is to measure it on different topological sectors. Simulations are on the way at the time of writing.


\subsection{Full QCD simulations}

In full QCD with two flavors of dynamical overlap fermions we concentrated on the channels $\pi, \delta, \eta', \sigma$ given by the correlators of the operators $\bar \psi \gamma_5 \vec \tau \psi, \bar \psi \vec \tau \psi,\bar \psi \gamma_5 \psi,\bar \psi \psi$, respectively. Here, $\tau$ is the Pauli matrix in the flavor space. 

If chiral symmetry is restored we expect that the pairs $(\sigma, \pi)$, $(\eta', \delta)$ become degenerate. Similarly, if the flavour-singlet axial symmetry is restored too at high temperatures we would have that all the channels become degenerate. The $\pi$ and the $\eta'$ differ just by the disconnected part, which is essentially given by the near-zero modes. This observation implies that the $U(1)_A$ breaking is driven by near-zero modes. By looking at the spatial correlators in those channels, we can check whether it happens. The problem of establishing if the axial symmetry is restored at the critical point is still open and has a relevance on the possible order of the phase transition \cite{Butti:2003nu}.

\begin{figure}[t]
\center
\includegraphics[width=0.6\textwidth,clip=1]{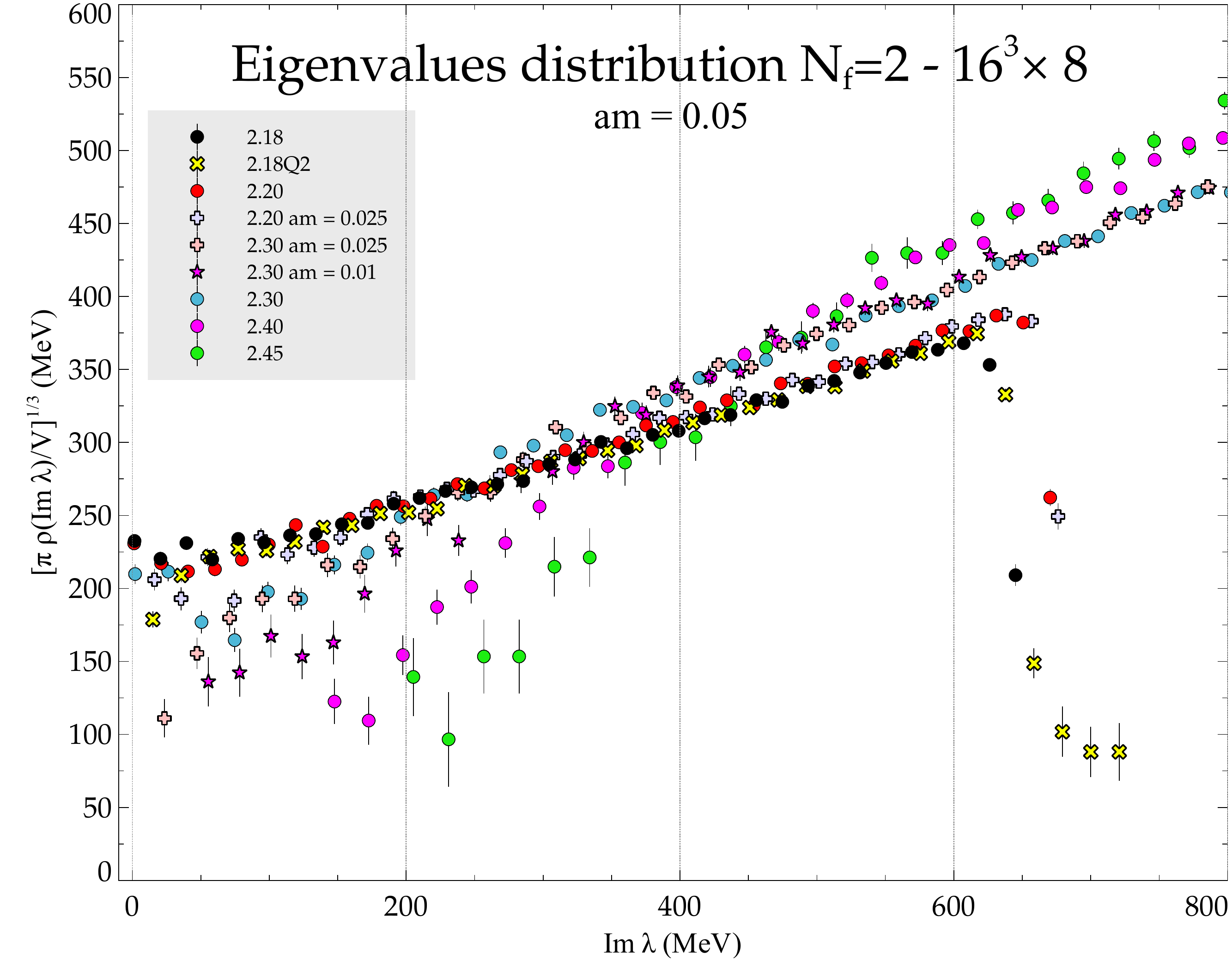}
\caption{Eigenvalues density in full QCD simulations, lattice $16^3\times 8$. Density around zero gives the chiral condensate by Banks-Casher relation.}
\label{fig:EigenvalNf2}
\end{figure}

Some details on the simulation follow. The algorithm is HMC, using Iwasaki action with topology fixing term and two flavors of sea quarks. The size of the lattice is $16^3\times 8$. We choose $N_t = 8$ to ensure that the configurations are smooth enough. Masses start from $am=0.05$ down to $am=0.01$ giving a pion mass of around 400 MeV at lowest temperature. The $\beta$s were chosen to be in the temperature range $T = [171, 243]$ MeV, $T/T_c = [0.97,1.39]$ (assuming the critical temperature to be $T_c = 175$ MeV). A comment is in order here: we couldn't estimate directly the transition temperature because it requires long history runs to measure susceptibilities. Anyway we checked that, by looking at eigenvalues of the Dirac operator (Figure \ref{fig:EigenvalNf2}), and the Polyakov loop, we simulated $\beta$s just below and above the transition temperature. The topological sector is mainly $Q=0$ but we have some simulations also at $Q=2$. All configurations were generated and analyzed using the BlueGene/L installation at KEK.

\begin{figure}[b]
\center
\includegraphics[width=0.48\textwidth,clip=1]{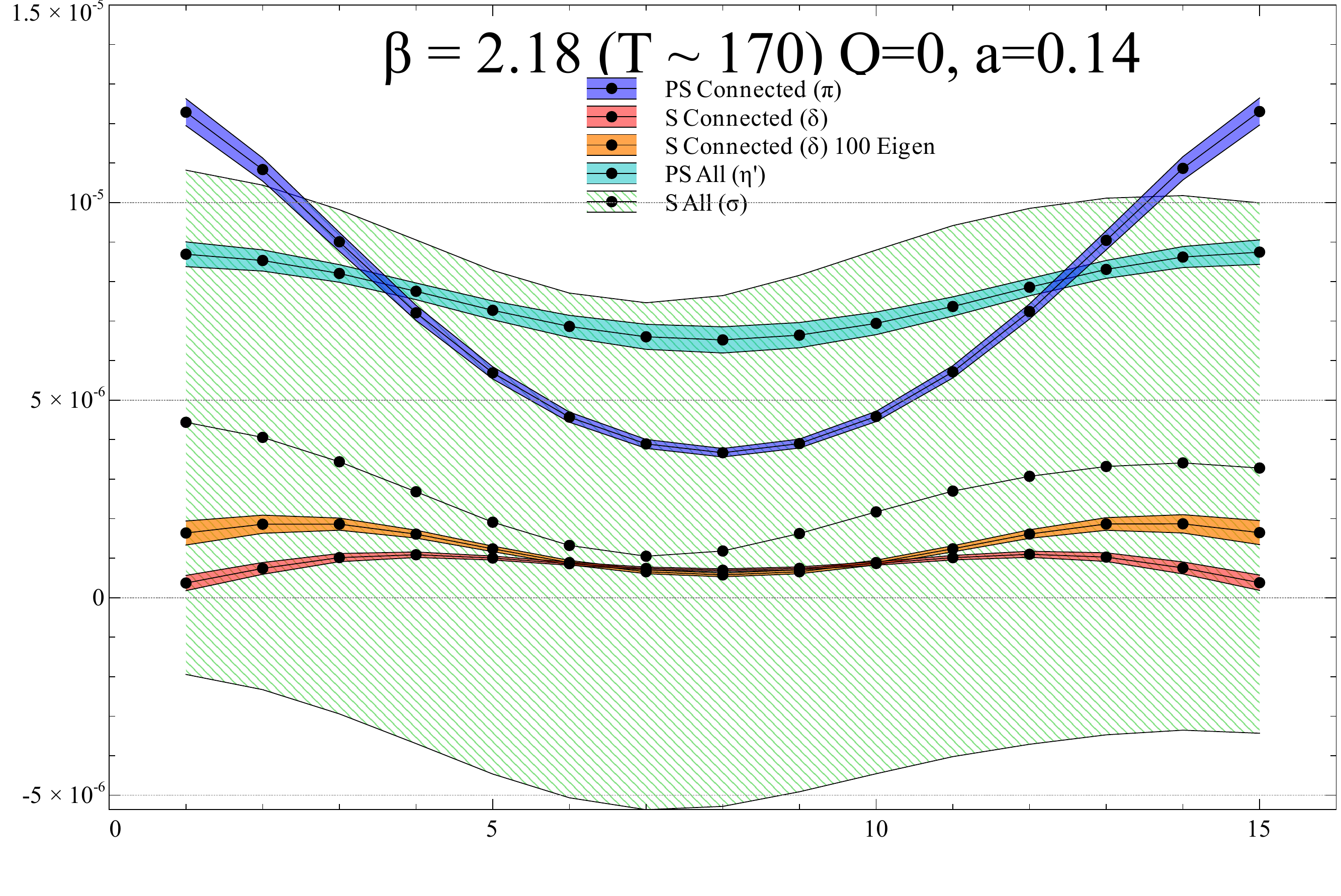}
\includegraphics[width=0.48\textwidth,clip=1]{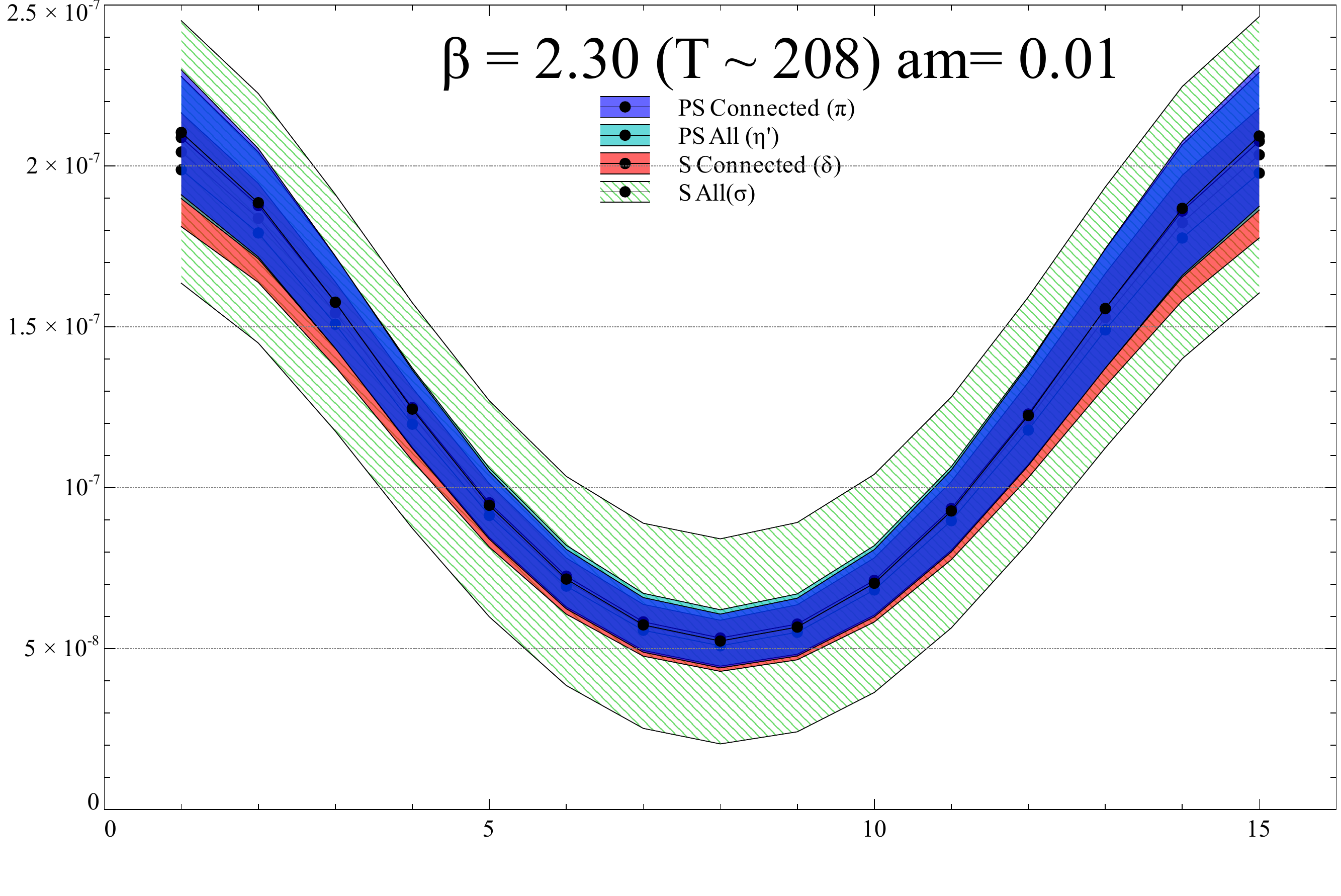}
\caption{Scalar and pseudoscalar spatial correlators at finite temperature. The estimate for the $\sigma$, green area, correlator has still huge errors in comparison to the others. We see some degeneracy for the correlators at $\beta = 2.30, T = 208$ MeV, right panel. } 
\label{fig:SPScorrelator}
\end{figure}

Some preliminary results are shown in Figure \ref{fig:SPScorrelator}. 
At this stage we cannot say anything quantitative since the volume seems still too small to extract masses. We can just check by inspection of the plots when the correlators become equal. We observe that the correlators start to become almost degenerate after the transition temperature and that as the sea quark mass is decreased this degeneracy is improved. We are collecting more data points to extrapolate this result toward the transition temperature.  

\section{Conclusions}

We described our project on finite temperature QCD with overlap fermions. Simulating overlap fermions forces to fix the topology in our case. A method to extract physical results from fixed $Q$ simulations was previosly developed for the zero temperature regime. To obtain reliable results we must check that we can apply the same method at finite temperature or find the necessary modifications. We started the investigation analyzing the behavior of topological susceptibility in pure gauge theory at $Q=0$. We found that our results differ at high temperature from the previous works. We are currently investigating the source of this discrepancy. A deep understanding of this problem is essential in the interpretation of full QCD results where we found restoration of axial symmetry at least from temperatures above $1.1T_c$. 

\section*{Acknowledgements}

Numerical simulations are performed on Hitachi SR11000 and IBM System Blue Gene Solution at High Energy Accelerator Research Organization (KEK) under a support of its Large Scale Simulation Program (No. 09/10-09). This work is supported in part by the Grant-in-Aid of the Ministry of Education, Culture, Sports, Science and Technology (No. 21674002, No. 20105005, No. 21684013, No. 220340047, No. 21105508) and by Grant-in-Aid for Scientific Research on Innovative Areas (No. 20105001, No. 20105003).

\end{document}